\documentclass[aip,reprint,pop,10pt,superscriptaddress,nobibnotes,showpacs]{revtex4-1}
\usepackage{amsfonts}
\usepackage{amsmath} 
\usepackage{amssymb} 
\usepackage{graphicx}%
\usepackage{color}

\begin{document}
\title{Large amplitude solitary waves in ion-beam plasmas with charged dust impurities}
\author{A. P. Misra}
\email{apmisra@visva-bharati.ac.in}
\affiliation{ Department of Mathematics, Siksha Bhavana, Visva-Bharati University, Santiniketan-731 235, India.}
\author{N. C. Adhikary}
\email{nirab$_$iasst@yahoo.co.in}
\affiliation{Material Sciences Division, Institute of Advanced Study in Science and Technology, Paschim Boragaon, Garchuk-781 035, Guwahati, Assam, India}

\begin{abstract}
The nonlinear propagation of large amplitude dust ion-acoustic (DIA) solitary waves (SWs) in an ion-beam  plasma with stationary charged  dusts is investigated. For  typical plasma parameters relevant for experiments [J. Plasma Phys. \textbf{60}, 69 (1998)], when the beam speed is larger than the DIA speed ($v_{b0}\gtrsim1.7c_s$), three stable waves, namely the  ``fast"  and  ``slow" ion-beam modes and the plasma DIA wave are shown to  exist. These modes  can propagate as SWs  in the beam plasmas. However, in the other regime ($c_s<v_{b0}<1.7c_s$), one of the beam modes when coupled to the DIA mode may become unstable.   The  SWs with positive (negative) potential may exist when the difference of the nonlinear wave speed ($M$) and the beam speed is such that $1.2\lesssim M-v_{b0}\lesssim1.6$ ($M-v_{b0}\gtrsim1.6$). Furthermore, for real density perturbations, the wave potential $(>0)$ is found to be  limited by a critical value which typically depends on $M$, $v_{b0}$   as well as the  ion/beam temperature. The  conditions for the existence of DIA solitons are obtained and their  properties  are analyzed numerically in terms of the system parameters. While the system  supports both the compressive and rarefactive large amplitude  SWs, the small amplitude solitons  exist only of the compressive type.   The theoretical results may be useful for observation of  soliton excitations in laboratory ion-beam driven plasmas  as well as in space plasmas where the charged dusts play as impurities.
\end{abstract}
\date{11 November 2011}
\maketitle
\section{Introduction}
Large amplitude solitary waves (SWs) in plasmas with a high energy ion beam have been observed at various space plasma environments, e.g., the Earth's magnetopause, in the Van Allen radiation belts as well as in the auroral zone \cite{SW1,SW2}. These nonlinear waves may be driven by the momentum exchange between the electrons and protons, or between different ion populations  in multi-ion plasmas. Propagation of such ion-acoustic (IA) SWs (IASWs) in magnetized or unmagnetized collisionless plasmas under different physical situations has been of considerable interest in recent years \cite{relativistic-ion-beam,ion-beam-amar,electron-beam-kappa,ion-beam-experiment-POP,dusty-negative-ion-PRE}.  Using different methods, many authors have studied the behavior and characteristics of IA solitons both theoretically \cite{relativistic-ion-beam,ion-beam-amar,ion-beam-theory-JPP,ion-beam-soliton,ion-beam-solitary,ion-beam-solitary-JGR} and experimentally \cite{ion-beam-experiment-POP,ion-beam-experiment-JPP,ion-beam-modes-PRL,beams-PRL} in ion-beam plasmas. 

It has been found that the presence of energetic charged particles like ion beam in plasmas can significantly modify the propagation behaviors of SWs \cite{particle-ion-beam}. The latter with negative potentials have been found in the vicinity of ion beam regions of the auroral zone in the upper atmosphere \cite{auroral-zone}. The spacecraft observations in the Earth's plasma sheet boundary  indicate that the electron and ion beams can also drive the broadband electrostatic waves \cite{broadband}.  Furthermore, the low-temperature plasmas containing charged dusts are ubiquitous in various space plasmas \cite{space},  laboratory devices  \cite{laboratory} as well as in industrial processes \cite{industry}. The presence of such negatively charged dusts can significantly influence the collective behaviors of plasmas \cite{DIA-theory,DIA-experiment}.

On the other hand, it has been shown  that when an ion beam is injected into an unmagnetized plasma, three stable normal modes, namely the ``fast" and ``slow" ion-beam modes and the plasma IA wave exist in plasmas \cite{ion-beam-modes-PRL,beams-PRL}. Furthermore, depending on whether the beam speed is of the order of or larger (e.g. 2 times) than the IA speed, the IA waves  when coupled to slow beam mode may become unstable. Such ion-ion instability  has been confirmed experimentally  by Gr\'{e}sillon  \textit{et al} \cite{ion-beam-modes-PRL}.  Thus, there may exist either three stable modes or one unstable and two stable modes.  The nonlinear wave evolution of such three  modes was investigated by Yajima \textit{et al} \cite{nonlinear-ion-beam} in an ion-beam plasma. They showed that each of these modes can propagate as small amplitude Korteweg-de Vries (KdV) solitons when the beam density is smaller than the electron density and the amplitude is smaller than a critical value. In an another work, Nakamura  \textit{et al} \cite{ion-beam-experiment-JPP} had experimentally shown that the propagation of large amplitude IASWs (compressive type) is possible for beam speed larger than the IA speed in an ion-beam plasma with two groups (high and low-temperature) of electrons. However, it seems that the effects of adiabatic positive ions and ion beam as well as their finite temperatures on the propagation of arbitrary amplitude dust IASWs (DIASWs) in a plasma with charged dust impurities, have not yet been considered in detail.  

In this paper, we study the properties of three  linear eigenmodes, which may propagate as stable DIASWs. For typical plasma parameters relevant for experimental conditions \cite{ion-beam-experiment-JPP}, one of the modes may become unstable when  the beam speed to dust ion-acoustic (DIA) speed ratio lies in $1<v_{b0}/c_s<1.7$.   The conditions for the existence of large amplitude DIASWs are analyzed by pseudopotential approach and their properties are studied numerically for fast modes. Depending on the parameters, we show the existence of   compressive as well rarefactive  DIASWs. The latter  exist when the ion to electron temperature ratio is of the order $\sim0.1$ or more. On the other hand,   the   DIASWs with small amplitudes are shown to propagate   only of the compressive type. 
\section{Basic equations and the dispersion relation}
We consider an unmagnetized collisionless   plasma  composed of adiabatic positive ions, positive  beam ions, Boltzmann distributed electrons and negatively charged (stationary) dust grains  in the background plasma. The finite temperatures of both the beam and  plasma ions together with an equilibrium flow of the beam are considered.  The normalized equations read
\begin{eqnarray}
& & \frac{\partial n_{j}}{\partial t}+\frac{\partial}{\partial x}\left(n_{j}v_{j}\right)=0,\label{1}\\
& & \frac{\partial v_{j}}{\partial t}+v_{j}\frac{\partial v_{j}}{\partial x}=-\frac{\partial \phi}{\partial x}-\frac{\sigma_{j}}{n_{j}}\frac{\partial P_{j}}{\partial x},\label{2}\\
& & \frac{\partial P_{j}}{\partial t}+v_{j}\frac{\partial P_{j}}{\partial x}+3P_{j}\frac{\partial v_{j}}{\partial x}=0,\label{3}\\
& & \frac{\partial^2 \phi}{\partial x^2}=\exp(\phi)+\mu_d-\mu_i n_i-\mu_b n_b, \label{4}
\end{eqnarray}
where $n_{j}$, $v_{j}$  and $P_{j}$ respectively denote the number density, speed and the pressure of singly charged beam $(j=b)$ and plasma ions $(j=i)$,  normalized by the  unperturbed number density $n_{j0}$, the ion-acoustic  speed  $c_s=\sqrt{k_BT_e/m_i}$ and $n_{j0} k_B T_{j}$. Here $k_B$ is the Boltzmann constant, $T_{j}$ is the temperature for $j$-th species ($j=e$ for electrons)  and $m_{j}$ is the particle mass with $m_i=m_b$. Furthermore,   $\phi$ is the DIA wave potential normalized by $k_BT_e/e$, where $e$ is the elementary charge. The space and time variables are respectively normalized by the Debye length $\lambda_D=\sqrt{k_BT_e/4\pi n_{i0}e^2}$ and the ion plasma period $\omega^{-1}_{pi}=1/\sqrt{4\pi n_{i0}e^2/m_i}$.  The overall charge neutrality condition in the background plasma reads
\begin{eqnarray}
\mu_i+\mu_b=1+\mu_d, \label{charge-neutrality}
\end{eqnarray}
where $\mu_{j}=n_{j0}/n_{e0}$ and $\mu_d=Z_dn_{d0}/n_{e0}$ in which $n_{d0}$ is the dust number density and $Z_d$ is the number of electrons  on the dust grain. In Eq. (\ref{2}), $\sigma_{j}=T_{j}/T_e$ is the temperature ratio for $j=i$, $b$. We have neglected the electron inertia, since the electron thermal speed is much larger than the ion/beam speed. 

It has been theoretically and experimentally shown that when beam ions are injected into an unmagnetized plasma, three longitudinal modes involving ion motions, namely an ion-acoustic wave and the fast and the slow space charge waves in the beam, can propagate \cite{ion-beam-modes-PRL,beams-PRL}. In order to identify those modes in our dusty beam plasma system we  linearize the basic equations and assume that the perturbations  vary as $\sim\exp(ikx-i\omega t)$, where $\omega$ is the angular frequency normalized by $\omega_{pi}$ and $k$  is the wave number normalized by $\lambda^{-1}_D$. Thus, we obtain the following dispersion relation.
\begin{eqnarray}
&& K^2=\frac{\mu_i}{\omega^2/k^2-3\sigma_i}+\frac{\mu_b}{\left(\omega/k-v_{b0}\right)^2-3\sigma_b}, \label{5-dispersion}
\end{eqnarray}
where $K^2=1+k^2$ and $v_{b0}$ is the speed (drift) of beam ions in equilibrium. In absence of the ion beam, DIA mode  propagates with the phase speed given by
\begin{eqnarray}
 \frac{\omega}k\approx\sqrt{3\sigma_i+\frac{\mu_i}{K^2}}. \label{IA}
\end{eqnarray}
 However, in absence of  plasma ions the fast (F) and  slow (S) modes propagate in the beam  with the phase speeds,
 \begin{eqnarray}
 \frac{\omega}k\approx v_{b0}\pm\sqrt{3\sigma_b+\frac{\mu_b}{K^2}}. \label{FS}
 \end{eqnarray} 
 \begin{figure*}
\includegraphics[width=7in,height=4in,trim=0.0in 0in 0in 0in]{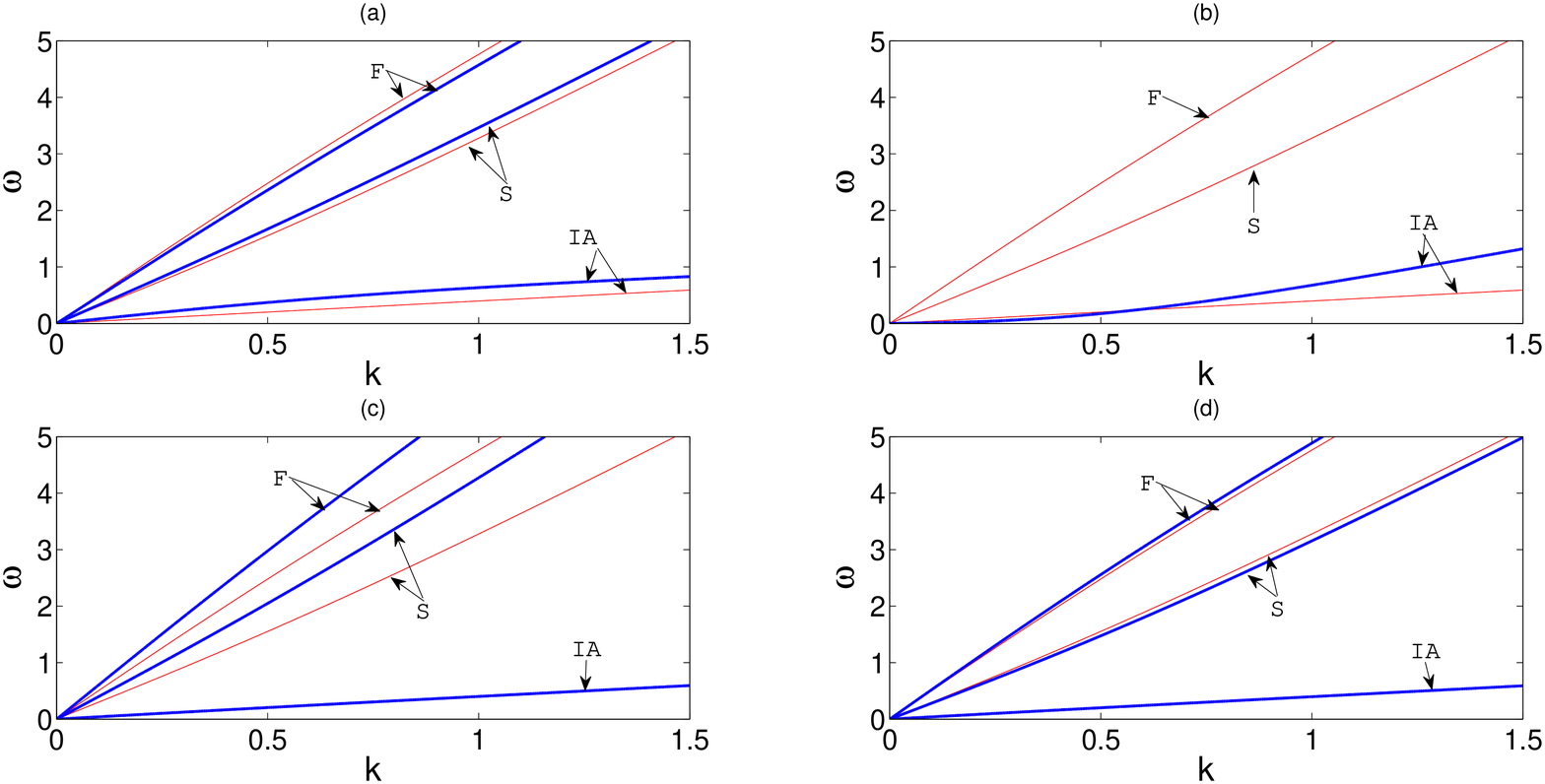} \caption{ (Color online) The dispersion equation (\ref{5-dispersion}) is contour plotted against the normalized wave number $(k)$ and the frequency $(\omega)$ for fast (F), slow (S) and DIA modes. The parameter values are (a) $v_{b0}=5$, $\sigma_{i}=0.05$, $\sigma_{b}=0.01$, $\mu_{d}=0.4$ with different $\mu_i$: $\mu_{i}=0.2$ [thin (red) line]  and $\mu_{i}=0.7$ [thick (blue) line]; (b) $v_{b0}=4$, $\mu_{i}=0.2$, $\sigma_{b}=0.01$, $\mu_{d}=0.4$ with different $\sigma_i$: $\sigma_{i}=0.05$ [thin (red) line]  and $\sigma_{i}=0.5$ [thick (blue) line]; (c) $\mu_{i}=0.2$, $\sigma_{i}=0.05$, $\sigma_{b}=0.01$, $\mu_{d}=0.4$ with different $v_{b0}$: $v_{b0}=4$ [thin (red) line]  and $v_{b0}=5$ [thick (blue) line] and $v_{b0}=4$, $\sigma_{i}=0.05$, $\sigma_{b}=0.01$, $\mu_{i}=0.2$ with different $\mu_d$: $\mu_{d}=0.4$ [thin (red) line]  and $\mu_{d}=0.8$ [thick (blue) line].}
\end{figure*}
{When the beam speed is much larger than the ion-acoustic speed, the beam and background ions are not strongly coupled. The background ions   support slightly modified ion-acoustic waves and the beam  ions support ion-acoustic waves (F and S modes) whose phase speeds are shifted by the beam speed $v_{b0}$. Equation \eqref{FS} shows that the phase speed of the fast (slow) modes  is  larger (smaller) than the equilibrium beam speed $v_{b0}$.} 
Now,  the dispersion equation (\ref{5-dispersion}) can be expressed as a polynomial equation in $\omega$ of degree $4$, and   when the  phase speed is much larger than the DIA speed, the dispersion relation  gives real roots of the wave frequency. However, in the opposite case, the coupled wave modes can be stable for
   $v_{b0}>v_{bc}$ where 
   \begin{eqnarray}
    v^2_{bc}\approx3(\sigma_i+\sigma_b)+\frac{\mu_i+\mu_b}{K^2}, \label{stable}
   \end{eqnarray}
   and unstable when
   \begin{eqnarray}
   3\sigma_b+\frac{3\sigma_i\mu_b}{\mu_i+3\sigma_i K^2}<v^2_{b0}<3(\sigma_i+\sigma_b)+\frac{\mu_i+\mu_b}{K^2}. \label{unstable}
   \end{eqnarray}
For typical laboratory plasma parameters \cite{ion-beam-experiment-JPP}  $\sigma_i=0.5$, $\sigma_b=0.01$, $\mu_i+\mu_b=1.4$  and $k=0.1$, the critical value of $v_{b0}$ above which the modes are stable is $v_{bc}\approx 1.7$,    and for unstable modes we have  $1<v_{b0}<1.7$. Thus, it is reasonable to consider higher values of the beam speed (i.e. roughly greater than 2 times the DIA speed) in order to avoid ion-ion instability \cite{ion-beam-experiment-JPP}. The latter has been confirmed in  experiments \cite{ion-beam-modes-PRL}.  

Next, we numerically examine the  dispersion relation (\ref{5-dispersion}) for the three wave modes which propagate along the beam direction. The results are shown   in Fig. 1 where a quartic equation in $\omega$ is contour plotted against the wave number. The behaviors of the modes are found similar as experimentally observed modes but in different beam plasmas \cite{ion-beam-experiment-JPP}.  It clearly shows how  the frequencies of the DIA wave as well as the F  and S modes get modified by the increase of the beam or ion density, the beam speed, the ion temperature as well as the percentage of impurity in the background plasma. We find that as  $\mu_i$ increases [Fig. 1(a)]  the phase speed of the F-modes (S-modes)  decreases (increases) and that for  the IA mode increases. However, the ion-temperature only modifies the DIA mode as in Fig. 1(b). Figure 1(c) shows that the effect of the beam speed is to enhance the phase speeds of both the F and S modes. In contrast to Fig. 1(a),   the enhancement of the charged dust concentration $\mu_d$ results into the increase (decrease) of the phase speed of F (S) modes, whereas the DIA modes remain almost unchanged by the effect of $\mu_d$.     In the following section we determine the conditions  for which these  stable modes would indeed  propagate as SWs, and   analyze their properties with different plasma parameters using the pseudopotential approach.
\section{Large amplitude solitons: Pseudopotential approach}
Assuming that the perturbations  vary in the moving frame of reference $\xi=x-Mt$, where $M$ is the  nonlinear wave speed  normalized by $c_s$ (If $M$ be normalized by the phase speed of the ion-acoustic waves, it would then be called as the Mach number), we obtain from Eqs. (\ref{1})-(\ref{4}) the following relation for the densities.
\begin{eqnarray}
&& n_{j}=\frac{\sigma_{j1}}{\sqrt{2}\sigma_{j0}}\left[\Phi_{j}-\sqrt{\Phi^2_{j}-a^2_{j}} \right]^{1/2},\label{6-density}
 \end{eqnarray}
 where $v_{i0}=0$, $\Phi_{j}=1-2\phi/\left(M-v_{j0}\right)^2\sigma^2_{j1}$, $\sigma_{j0}=\sqrt{3\sigma_{j}}/(M-v_{j0})$, $\sigma_{j1}=\sqrt{1+\sigma^2_{j0}}$ and $a_{j}=2\sigma_{j0}/\sigma^2_{j1}$. 
 In our nonlinear theory we will consider $M$ to be larger than the beam speed in order to examine whether the fast modes propagate as solitary waves \cite{ion-beam-experiment-JPP}.  Similar analysis can also be done for the slow modes with $M<v_{b0}$.
 In obtaining Eq. (\ref{6-density}) we have used the boundary conditions, namely $\phi\rightarrow0$, $v_{i,b}\rightarrow(0,v_{b0})$, $n_{j}\rightarrow1$ and $P_{j}\rightarrow1$ as $\xi\rightarrow\pm\infty$. 
 Furthermore, inspecting on Eq. (\ref{6-density}), we find that the number densities are real when there exists a critical value $\phi_c$ of $\phi$ such that $0<\phi<\phi_{c}$  where
\begin{eqnarray} 
\phi_{c}=\min\left\lbrace\frac{M^2}{2}\left(1-\sigma_{i0}\right)^2,\hskip1pt \frac12{\left(M-v_{b0}\right)^2}\left(1-\sigma_{b0}\right)^2\right\rbrace.\nonumber\\ \label{phi-max}
\end{eqnarray}
Note that the above restriction is valid only for large amplitude  waves with positive potential. However, for waves with negative potential, the values of $\phi$ are not limited by the ratios, namely $\mu_i$, $\mu_b$ or $\sigma_i$, $\sigma_b$. 
 Typically, for $M>v_{b0}$ and $\sigma_b<\sigma_i<1$, $\phi_c$ assumes the second term in the curly brackets in Eq. (\ref{phi-max}). However, this value must be considered together with the conditions for the existence of SWs.    Introducing now the relation (\ref{6-density}) into Eq. (\ref{4}),  and integrating it we obtain the following energy balance equation for an oscillating particle of unit mass at the pseudoposition $\phi$ and pseudotime $\xi$.
 \begin{eqnarray}
 && \frac{1}{2}\left(\frac{d\phi}{d\xi}\right)^2+V(\phi)=0, \label{7-energy}
  \end{eqnarray}
where  the pseudopotential $V$ is given by
 \begin{align}
 V(\phi)&=1-\exp{(\phi)}-\mu_d\phi-\frac{1}{3}\sum\limits_{j=i,b}\beta_j\nonumber\\
 &\times\left[\left(\Phi_j-\sqrt{\Phi^2_j-a^2_j}\right)^{3/2}-\left(1-\sqrt{1-a^2_j}\right)^{3/2}+3a^2_j\right.\nonumber\\
 &\left.\times\left\lbrace\left(\Phi_j-\sqrt{\Phi^2_j-a^2_j}\right)^{-1/2}-\left(1-\sqrt{1-a^2_j}\right)^{-1/2}\right\rbrace\right].\label{8-V} 
  \end{align}
Here $\beta_j=\mu_j(M-v_{j0})^2\sigma^3_{j1}/2\sqrt{2}\sigma_{j0}$. Equations (\ref{7-energy}) and (\ref{8-V}) are valid for arbitrary amplitude  stationary   perturbations like SWs and/or double layers. The  conditions for the existence of such perturbations  can be obtained as follows:

(i) $V(0)=0$. This has already been satisfied in obtaining Eq. (\ref{8-V}),  and by using the charge neutrality condition (\ref{charge-neutrality}), one can  easily verify that $dV(\phi)/d\phi=0$ at $\phi=0$;

(ii) $d^2V(\phi)/d\phi^2<0$ at $\phi=0$. This is satisfied when the following inequality holds.
\begin{eqnarray}
\frac{\mu_i}{M^2-3\sigma_i}+\frac{\mu_b}{(M-v_{b0})^2-3\sigma_b}<1. \label{derivative-condition}
\end{eqnarray}

(iii) $V(\phi_m\neq0)=0$ and $dV(\phi_m)/d\phi\gtrless0$ according to whether the SWs are compressive (with positive potential, i.e.  $\phi>0$) or rarefactive (with negative potential, i.e. $\phi<0$). Here $\phi_m$ represents the amplitude of the solitary waves or double layers, if exist.

In order that the three modes propagate as  SWs  the conditions (i)-(iii) must be satisfied. However, for the double layers to exist  there must be an additional condition, i.e., $dV(\phi_m)/d\phi=0$  to be satisfied along with (i)-(iii). 
 We numerically analyze the conditions (ii) and (iii) for some typical plasma parameters as relevant for experiments \cite{ion-beam-experiment-JPP}.   The condition (ii), which results to Eq. (\ref{derivative-condition}) is presented in the $\mu_i-M$ plane as shown in Fig. 2 for different sets of parameters. The white (gray or shaded) regions show the parameter space where the inequality  (\ref{derivative-condition}) is  satisfied (not satisfied).  
The upper boundary curves  that separate the white and gray regions  represent the minimum values of $M$ $(>v_{b0})$ for the existence of solitary waves or double layers. The upper limits of $M$ (with $M>v_{b0}$ and  $\sigma_b<\sigma_i<1$) for the existence of positive SWs can be obtained by substituting $\phi=\phi_c\equiv\frac12\left(M-v_{b0}\right)^2\left(1-\sigma_{b0}\right)^2$ into $V(\phi)=0$ [cf. Eq. (\ref{8-V})]. These limiting values of $M$ are represented  by the thick (blue) curves in Fig. 2.   Typically, for $M>v_{b0}$, Fig. 2 shows that  the lower and upper  limits of $M$ increase with  the  beam speed as well as the charged dust concentration. However, the case of ion temperature effect [\textit{c.f.} Figs. 2(a) and (c)] is different. Here the  range of $M$ ($>v_{b0}$) remains almost the same with the increase of $\sigma_i$.  Thus, for $M>v_{b0}$, SWs may exist in the upper parts of the white regions. 
\begin{figure*}
\includegraphics[width=7in,height=4in,trim=0.0in 0in 0in 0in]{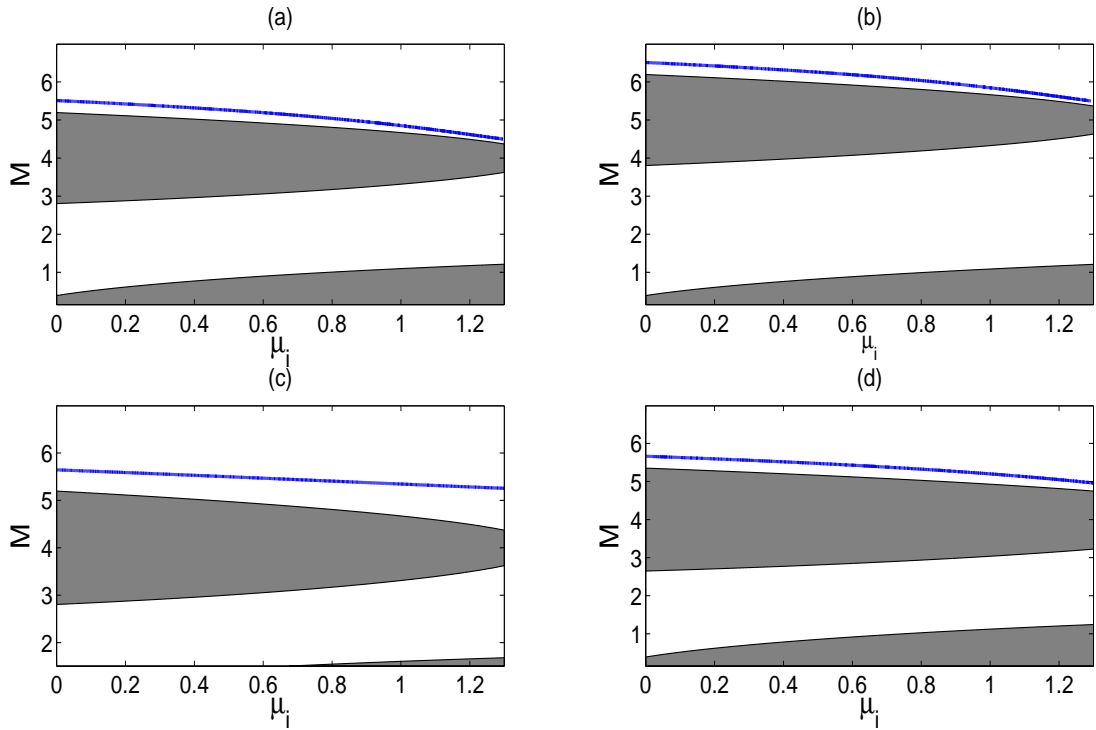} \caption{ The parametric regions (white or blank) are shown for $d^2V/d\phi^2<0$ at $\phi=0$ in the $(M-\mu_i)$ planes for (a) $v_{b0}=4$, $\mu_{d}=0.4$, $\sigma_i=0.05$ and $\sigma_{b}=0.01$; (b) the same as in (a) but $v_{b0}=5$; (c) the same as in (a) but $\sigma_{i}=0.5$; and (d) the same as in (a) but $\mu_{d}=0.8$. The shaded (gray) regions correspond to $d^2V/d\phi^2>0$ at $\phi=0$. Thus, for $M>v_{b0}$, SWs may exist only in the upper parts of the white regions. The thick (blue) curves represent the upper limits of $M$ for the existence of positive SWs.} 
\end{figure*}
\begin{figure*}
\includegraphics[width=7in,height=4in,trim=0.0in 0in 0in 0in]{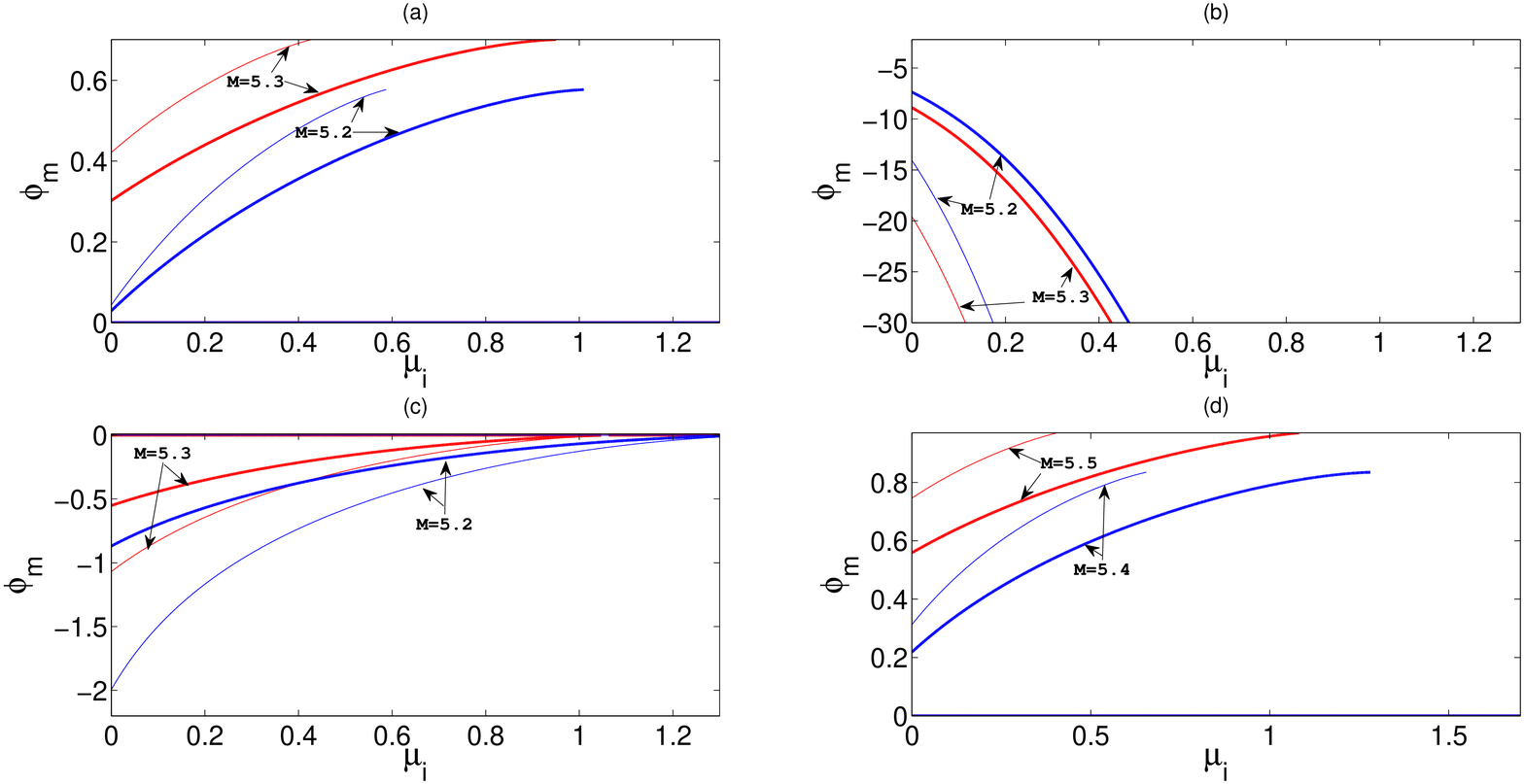} \caption{ (Color online) The contour lines are shown along which $V(\phi_m)=0$ (thin lines) and $dV(\phi_m)/d\phi=0$ (thick lines) for different $M$ as in the figure. The other parameter values for the subplots (a), [(b) and (c)] and (d) are the same as in Fig. 2(a), Fig. 2(c) and Fig. 2(d) respectively. The   values of $\phi=\phi_m$ are such that $0<\phi_m<\phi_{c}$  for positive  SWs, where $\phi_c$ is given by Eq. \eqref{phi-max}. The subplots (b) and (c) are shown in two different regimes of $\phi_m$ where the formation of rarefactive (negative) SWs may be possible. The plots (a) and (d) show the regions for the  existence of compressive SWs. In all the plots, there is no common point of intersection of  $V(\phi_m)=0$  and $dV(\phi_m)/d\phi=0$ ($\phi_m\neq0$) implying that the formation of double layers is not possible.  The common regions  where $\phi_m>0$  and $dV(\phi_m)/d\phi>0$ [left to or  above the curves in plots (a) and (d)] are for positive SWs and those where $\phi_m<0$  and $dV(\phi_m)/d\phi<0$ [left to or below  the curves in plots (b) and (c)] are for negative SWs.  }
\end{figure*}

Figure 3 shows  the contour lines of  $V(\phi)=0$ [thin (red) curves] and  $dV(\phi)/d\phi=0$ [thick (blue) curves] for different values of $M$. There are, in general, wide ranges of  values of  $\mu_i$ and $\phi_m$, for which $V=0$ and $dV/d\phi=0$ can be satisfied.  However, all the numerical values of $\phi=\phi_m$ are not admissible  as values of $\phi=\phi_m$ is limited by $\phi_c$ such that  $0<\phi_m<\phi_{c}$  [cf. Eq. \eqref{phi-max}].  Such $\phi_c$ changes with $M$ or some other parameters $\sigma_i$, $\sigma_b$. Thus, for each set of parameters we can obtain a value of $\phi_c$ below (for positive $\phi$)  which  SWs may exist.   Figures 3(a) and (d) explain the variation of the charged dust impurity and Figs. 3(b) and (c) that due to  ion temperature  in different (higher and lower) regimes of $\phi_m$. We find that the enhancement of the  ion temperature ($0.1\lesssim\sigma_i\lesssim0.5$) can lead to the existence of negative  SWs [Figs. 3(b) and (c)]. The common parameter regimes (left to or above the curves) in  Figures 3(a) and (d)   where both the conditions $\phi>0$ and $dV/d\phi>0$ are satisfied gives the region of the existence of positive SWs, whereas the conditions  $\phi<0$ and $dV/d\phi<0$ are satisfied only to the left to or down  the curves in Figs. 3(b) and (c), implying the existence of negative SWs.   Note, however, that with similar parameter values as in Fig. 3(b) or (c), the formation of negative SWs was not reported in beam-plasma experiments \cite{ion-beam-experiment-JPP}.  Since there is no common point of intersection of  $V(\phi)=0=dV(\phi)/d\phi$ except at $\phi=0$, we  conclude that the double layer formation may not be possible in our system. 
\begin{figure*}
\includegraphics[width=7in,height=4in,trim=0.0in 0in 0in 0in]{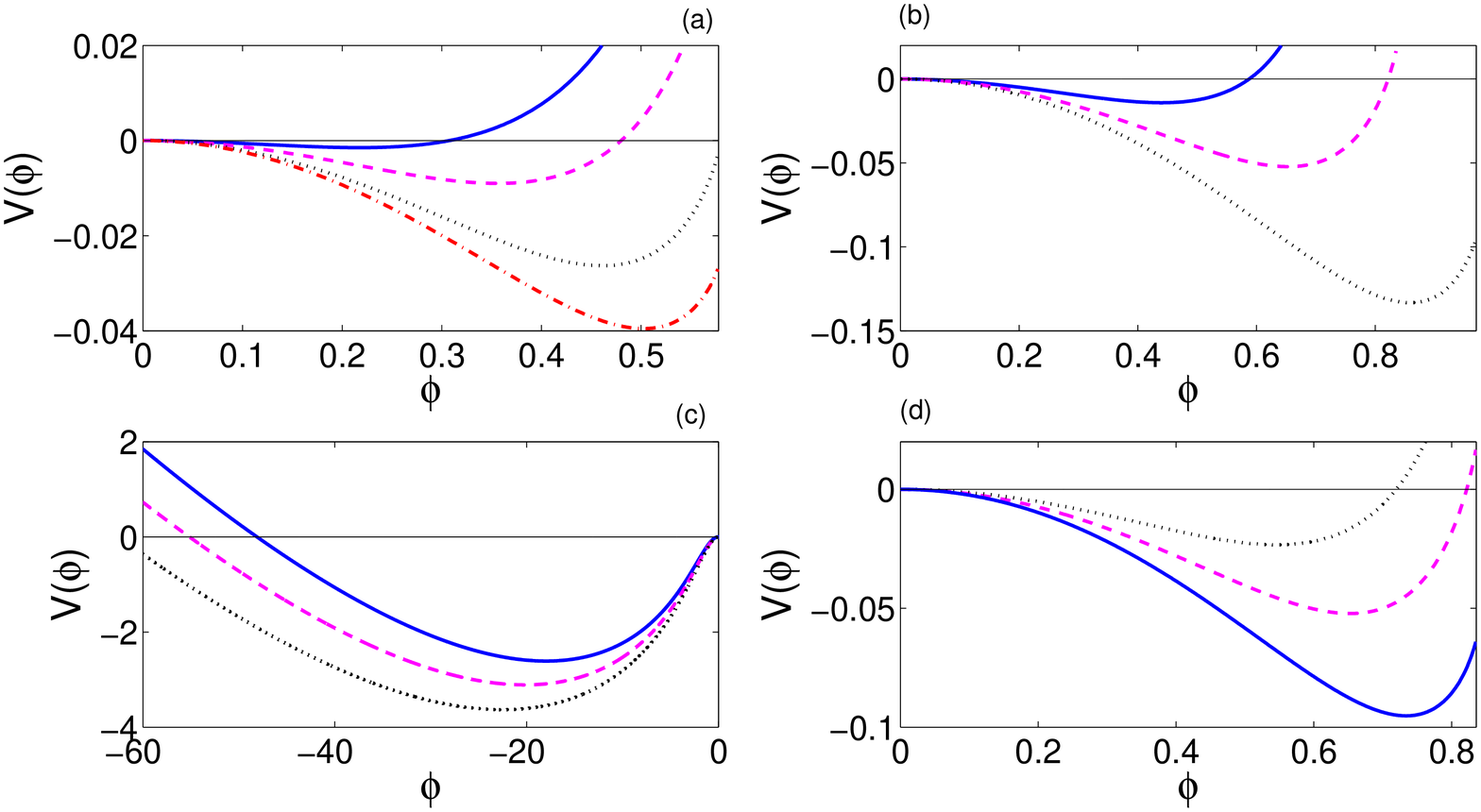} \caption{ (Color online) The pseudopotential $V(\phi)$ is plotted against $\phi$ to show the existence of positive [subplots (a), (b) and (d)] and negative [subplot (c)]  SWs. Subplot (a): The solid, dashed, dotted and dash-dotted lines  respectively correspond to $\mu_i=0.2$, $0.4$, $0.6$ and $0.7$ for a fixed $M=5.2$. The other parameters are as in Fig. 2(a). Subplot (b): The parameter values are as in subplot (a)  with $\mu_i=0.2$, and for $M=5.3$ (solid line), $5.4$ (dashed line) and $5.5$ (dotted line). Subplot (c): The parameter values are as in Fig. 2(c)  with $\mu_i=0.1$, and for $M=5.6$ (solid line), $5.7$ (dashed line) and $5.8$ (dotted line). Subplot (d): The solid, dashed and dotted lines represent respectively for $\mu_d=0.2$, $0.4$ and $0.6$. The parameter values are $\mu_i=0.2$, $M=5.4$ and others as in Fig. 2(a). The points where $V$ crosses the $\phi$-axis represent the amplitude of the SWs and the widths can be obtained as $|\phi_m|/|V_{\min}|$. }
\end{figure*}

Next, the pseudopotential $V(\phi)$ can be plotted against $\phi$ for a given set of physical parameters. We see from Fig. 4 that for both positive and negative values of $\phi$, $V(\phi)$ crosses the $\phi$-axis for the fast modes, implying the existence of both positive and negative SWs. These different crossing  points for certain values of the parameters  give the height or amplitude $|\phi_m|$ of the SWs. However, the width  of the SWs with speed $M$ can be obtained from the shape of $\phi(\xi)$ by the numerical solution of Eq. (\ref{7-energy}).   We can also verify the wave amplitude and width [$|\phi_m|/|V_{\min}|$, where $|V_{\min}|$ is the minimum value of $V(\phi)$]  obtained from the plots of $V(\phi)$ with those from the  numerical  solution of Eq. (\ref{7-energy}).     Since the  values of $\phi$ are limited by $\phi_c$, i.e., $0<\phi<\phi_{c}$ , and $\phi_c$ changes with $M$ and other parameters, $V(\phi)$ can  cross the $\phi$-axis only for a set of parameters. We find that the amplitude of the wave increases with the enhancement of the ion concentration until $\mu_i<0.6$ [Fig. 4(a)] as well as the wave speed $M\lesssim5.4$ [Fig. 4(b)], and it decreases with the increase of the dust impurity,   $\mu_d\gtrsim0.4$ [Fig. 4(d)]. On the other hand, Fig. 4(c) shows  the existence of negative SWs with large negative values of $\phi$.
The amplitudes (in absolute value) of these SWs increase with the nonlinear wave speed $M$. For the parameters as in Fig. 4(c) we find that in contrast to positive SWs in which upper limits of $M$ exist, there exist a lower limit of $M$ $(\gtrsim5.6)$ above which the SWs with negative potential exist.
\begin{figure*}
\includegraphics[width=5in,height=3.0in,trim=0.0in 0in 0in 0in]{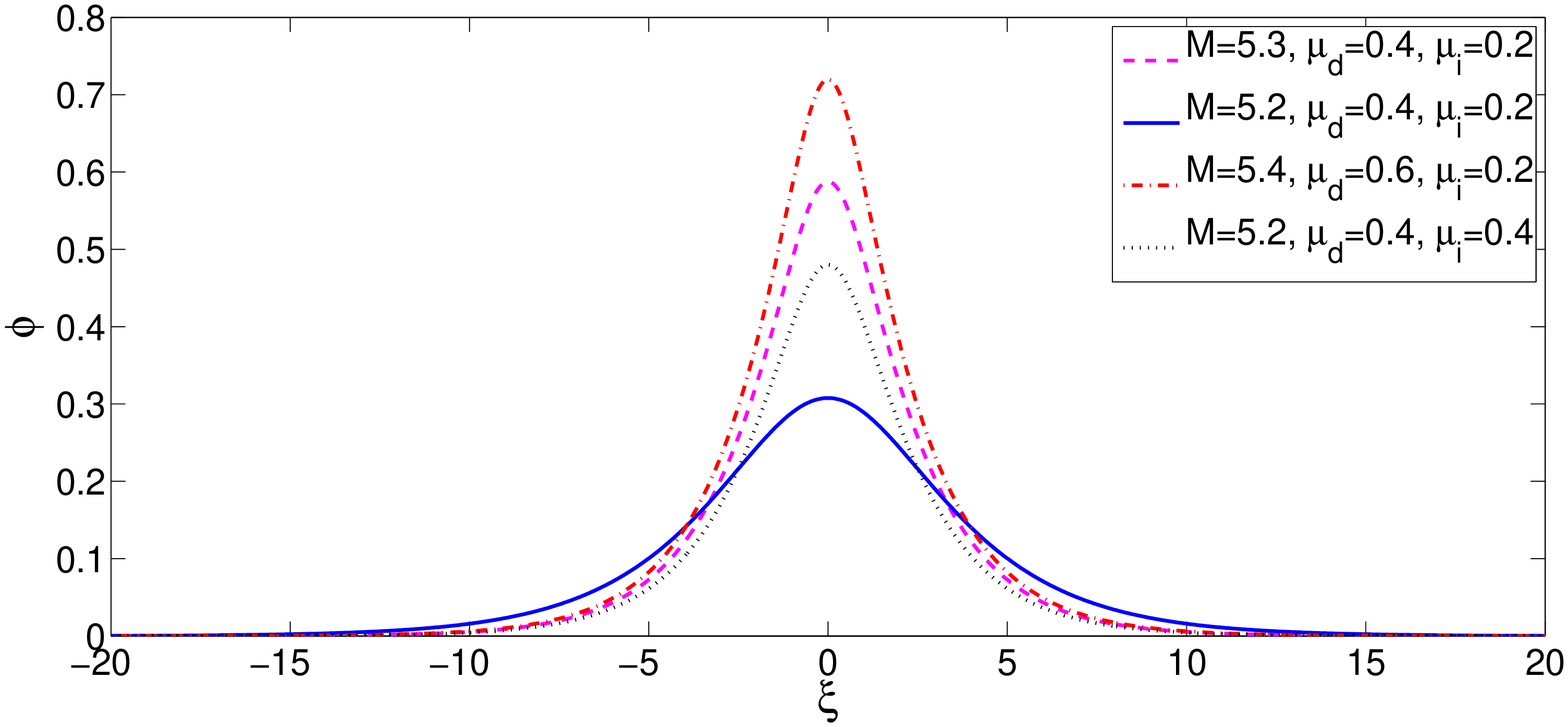} \caption{ (Color online) Numerical soliton solution (compressive) of Eq. (\ref{7-energy}) for the parameters as indicated in the figure. The other parameters are as in Fig. 4(b). }
\end{figure*}
\begin{figure*}
\includegraphics[width=5in,height=3.0in,trim=0.0in 0in 0in 0in]{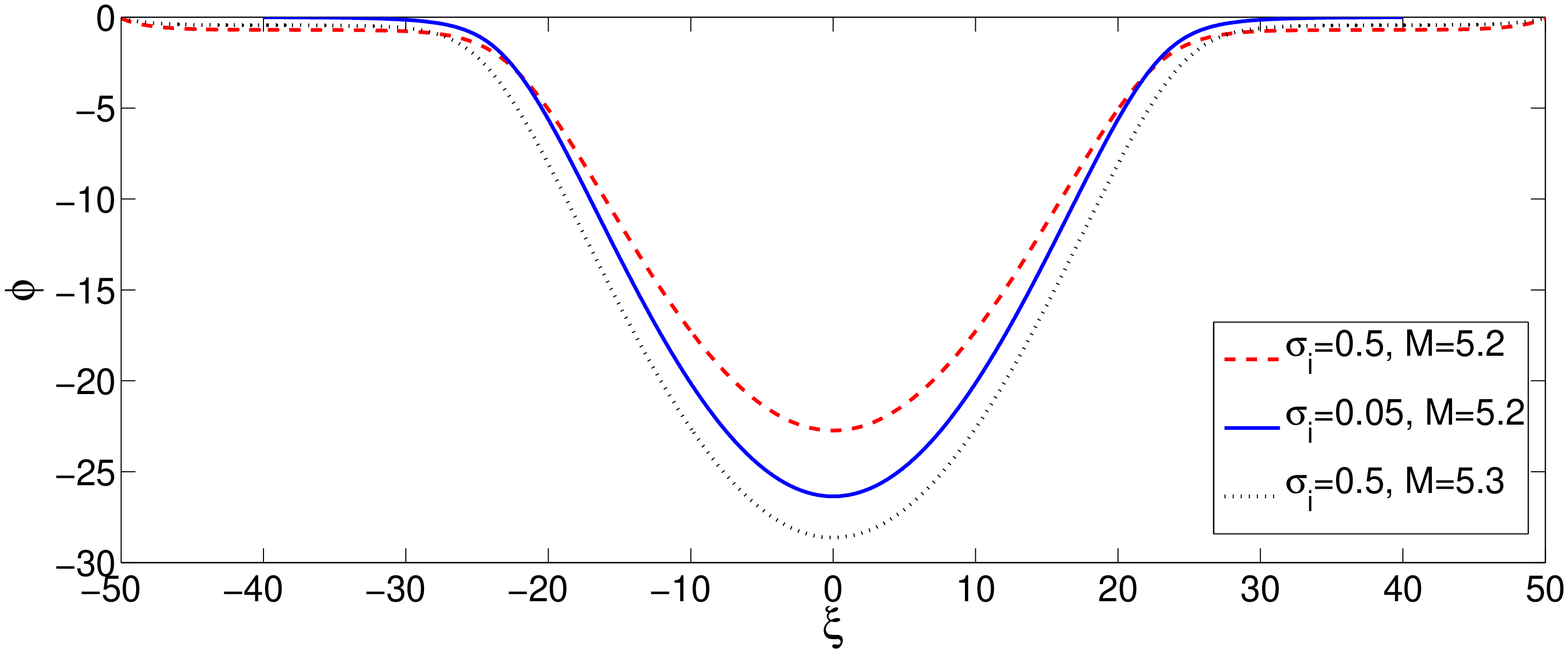} \caption{ (Color online) Numerical soliton solution (rarefactive) of Eq. (\ref{7-energy}) for $M=5.2$ (dashed line) and $M=5.3$ (dotted line). The other parameters are as in Fig. 4(c).  }
\end{figure*}

To verify the results as in Fig. 4 (especially the amplitude and width of the SWs) and to obtain the shape of $\phi(\xi)$, we numerically integrate Eq. (\ref{7-energy}). The results are presented in Figs. 5 and 6 for the positive and negative SWs respectively. We find that for compressive SWs (Fig. 5), the amplitude increases and the width decreases with increasing  values of  $M$ and $\mu_i$, whereas   both the amplitude and the width are enhanced by the effect of charged dust impurity $\mu_d$. Figure 6 shows that  the effects of $M$ and $\sigma_i$ respectively are to increase and decrease the amplitude as well as the width of the negative SWs.
\section{Summary and Conclusion} 
The nonlinear evolution of large amplitude SWs in a dust contaminated ion-beam driven plasma  is investigated by using a pseudopotential approach. The dispersion properties of three linear modes, namely the fast and slow ion-beam modes and the DIA wave are analyzed numerically.  The conditions for which these three modes propagate as  (large amplitude) SWs and their (soliton) properties are  studied numerically in terms of the system parameters.  
While the system supports both compressive (positive) and rarefactive (negative) large amplitude DIASWs, the small amplitude DIA solitons exist only of the compressive type (see Fig. 5).    The presence of charged dusts significantly alters the existence regions as well as the properties of solitons. We also show that for typical plasma parameters as in experiments \cite{ion-beam-experiment-JPP} the formation of double layers is not possible.  Our results can be summarized as follows: 

(i) When the beam speed is larger than ($\sim2$ times) the DIA speed, each of the stable modes, namely the fast and slow beam mode as well as the DIA mode can propagate as SWs  in ion-beam plasmas. Otherwise, there may exist one  unstable beam mode coupled to the DIA wave and two stable modes in the plasma.  With the enhancement of $\mu_i$, while the phase speed of the F-modes (S-modes)  decreases (increases)  that for  the DIA mode increases. These behaviors are in contrast to the effect of charged dusts  $\mu_d$ in which the DIA modes remain almost unchanged.   The effect of the beam speed is to enhance the phase speed of both the F and S-modes. In the limit of the phase speed larger than the DIA speed, the modes propagate as stable waves. However, in the opposite limit, the DIA mode,  coupled to a beam mode, may become unstable in $1<v_{b0}<1.7$ for typical plasma parameters $\sigma_i=0.5$, $\sigma_b=0.01$, $\mu_i+\mu_b=1.4$  at $k=0.1$, as relevant for experiments \cite{ion-beam-theory-JPP}. Thus, it is reasonable to consider higher values of $v_{b0}$ ($\gtrsim2$) for the DIASWs to exist.

(ii)  For the perturbations to be real, the wave potential $\phi$ $(>0)$ is to be less than a critical value, which typically depends on the nonlinear wave speed, the beam speed  as well as the beam or the ion temperature. 

(iii) Both the compressive and rarefactive large amplitude DIASWs may coexist, whereas the small amplitude soliton exists only of the compressive type. In order that the the  SWs with $\phi>0$ ($\phi<0$) may exist, the regime of the nonlinear wave speed ($M$) and the beam speed ($v_{b0}$) is such that $1.2\lesssim M-v_{b0}\lesssim1.6$ $(M-v_{b0}\gtrsim1.6)$.

(iv) For large amplitude positive SWs, the effects of  $M$ and $\mu_i$  are to enhance the wave amplitude and to reduce the width. However,  both the amplitude and the width may be increased (in magnitude) by the charged dust concentration $\mu_d$. These behaviors are similar to the case of large amplitude negative SWs by the effect of $M$. 

The  theoretical results could be useful  for soliton excitation in laboratory ion-beam driven plasmas as well as in space plasmas where negatively charged dusts are considered as  impurities. 

\acknowledgments{Authors sincerely thank the Referee for his useful comments which improved the manuscript in the present form.  APM was supported by the Kempe Foundations, Sweden through Grant No. SMK-2647.}


\begin{thebibliography}{150}
\bibitem{SW1} C. Cattell, J. Crumley, J. Dombeck, J. Wygant, and F. S. Mozer, Geophys. Res. Lett. \textbf{29}, 1807 (2002).
\bibitem{SW2} R. E. Ergun, C. W. Carlson, J. P. McFadden, F. S. Mozer, G. T. Delory, W. Peria, C. C. Chaston, M. Temerin, I. Roth, L. Muschietti, R. Elphic, R. Strangeway, R. Pfaff, C. A. Cattell, D. Klumpar, E. Shelley, W. Peterson, E. Moebius, and L. Kistler, Geophys. Res. Lett. \textbf{25}, 2041 (1998).
\bibitem{relativistic-ion-beam} B. C. Kalita, R. Das, and H. K. Sarmah, Phys. Plasmas \textbf{18}, 012304 (2011).
\bibitem{ion-beam-amar} N. C. Adhikary A. P. Misra, H.Bailung, and J. Chutia,  Phys. Plasmas \textbf{17}, 044502 (2010).
\bibitem{electron-beam-kappa} N. S. Saini and I. Kourakis, Plasma Phys. Control. Fusion \textbf{52}, 075009 (2010).
\bibitem{ion-beam-experiment-POP} S. K. Sharma and H. Bailung, Phys. Plasmas \textbf{17}, 032301 (2010).
\bibitem{dusty-negative-ion-PRE} A. A. Mamun, B. Eliasson, and P. K. Shukla,   Phys. Rev. E \textbf{80}, 046406 (2009).
\bibitem{ion-beam-theory-JPP} Y. Nakamura and K. Ohtani, J. Plasma Phys.  \textbf{53}, 235 (1995).
\bibitem{ion-beam-soliton} P. S. Abrol and S. G. Tagare,  Plasma Phys. \textbf{22}, 831 (1980); Phys. Lett. A \textbf{75}, 74 (1979) .
\bibitem{ion-beam-solitary} B. Karmakar, G. C. Das, and Kh. I. Singh,  Plasma Phys. Control. Fusion \textbf{30}, 1167 (1988).
\bibitem{ion-beam-solitary-JGR} V. A. Marchenko and M. K. Hudson, J. Geophys. Res. \textbf{100}, 19791 (1995).
\bibitem{ion-beam-experiment-JPP}   Y. Nakamura and K. Komatsuda, J. Plasma Phys.  \textbf{60}, 69 (1998).
\bibitem{ion-beam-modes-PRL} D. Gr\'{e}sillon and F. Doveil,  Phys. Rev. Lett. \textbf{34}, 77 (1975).
\bibitem{beams-PRL} T. Honzawa,  Phys. Rev. Lett. \textbf{53}, 1915 (1984).
\bibitem{particle-ion-beam} E. Okutsu, M. Nakamura, Y. Nakamura, and T. Itoh, Plasma Phys. \textbf{20}, 561 (1978).
\bibitem{auroral-zone} M. Temerin, K. Cerny, W. Lotko, and F. S. Mozer, Phys. Rev. Lett. \textbf{48}, 1175 (1982).
\bibitem{broadband} G. Parks, L. J. Chen, M. McCarthy, D. Larson, R. P. Lin, T. Phan, H. Reme, and T. Sanderson, Geophys. Res. Lett. \textbf{25}, 3285 (1998).
\bibitem{space} E. C. Whipple, T. G. Northrop, and D. A. Mendis, J. Geophys. Res. \textbf{90}, 7405 (1985).
\bibitem{laboratory} D. P. Sheehan, M. Carilo, and W. Heidbrink, Rev. Sci. Instrum. \textbf{61}, 3871 (1990).
\bibitem{industry} G. S. Selwyn, J. Singh, and R. S. Bennett,  J. Vac. Sci. Technol. \textbf{A7},  2758 (1989).
\bibitem{DIA-theory} P. K. Shukla and V. P. Silin, Phys. Scr. \textbf{45}, 508 (1992).
\bibitem{DIA-experiment} A. Barkan, R. L. Merlino, and N. D'Angelo, Phys. Plasmas \textbf{2}, 3563 (1995).
\bibitem{nonlinear-ion-beam} N. Yajima, M. Koni, and S. Ueda, J. Phys. Soc. Japan \textbf{52}, 3414 (1983). 


\end{thebibliography}
\end{document}